# Mixed Valence Driven Heavy-Fermion Behavior and Superconductivity in $KNi_2Se_2$

James R. Neilson,[1,2,†] Anna Llobet,[3] Andreas V. Stier,[2] Liang Wu,[2] Jiajia Wen,[2] Jing Tao,[4] Yimei Zhu,[4] Zlatko B. Tesanovic,[2] N. P. Armitage,[2] and Tyrel M. McQueen[1,2,*]

[1] Department of Chemistry, Johns Hopkins University, Baltimore MD.
[2] Institute for Quantum Matter, and Department of Physics and Astronomy, Johns Hopkins University, Baltimore MD.
[3] Lujan Neutron Scattering Center, Los Alamos Neutron Science Center, Los Alamos National Laboratory, Los Alamos, NM.
[4] Condensed Matter Physics and Materials Science Department, Brookhaven National Laboratory, Upton, NY.

[†] jneilso2@jhu.edu
[*] mcqueen@jhu.edu

Based on specific heat and magnetoresistance measurements, we report that a "heavy" electronic state exists below $T \approx 20$ K in $KNi_2Se_2$, with an increased carrier mobility and enhanced effective electronic band mass, $m^* = 6m_b$ to $18m_b$. This "heavy" state evolves into superconductivity at $T_c = 0.80(1)$ K. These properties resemble that of a many-body heavy-fermion state, which derives from the hybridization between localized magnetic states and conduction electrons. Yet, no evidence for localized magnetism or magnetic order is found in $KNi_2Se_2$ from magnetization measurements or neutron diffraction. Instead, neutron pair-distribution-function analysis reveals the presence of local charge-density-wave distortions that disappear on cooling, an effect opposite to what is typically observed, suggesting that the low-temperature electronic state of $KNi_2Se_2$ arises from cooperative Coulomb interactions and proximity to, but avoidance of, charge order.

## Introduction

Many-body states, in which complex phenomena arise from simple interactions, often exhibit rich phase diagrams and lack formal predictive theories. The emergence of exotic many-body quantum ground states in solids frequently occurs in proximity to a long range ordered state (*e.g.*, antiferromagnetism), as in high-temperature superconductivity[1,2], valence bond condensation in frustrated magnets[3], and the many-body heavy-fermion state[4-6]. Mixed valency[7], in which a compound has an element in more than one formal oxidation state, is responsible for a spectacular array of properties, from the catalytic power of many enzymes,[8] to the Verwey transition in magnetite[9,10]. In one case, suppression of disproportionation and charge-order gives rise to the emergence high-temperature superconductivity in $Ba_{1-x}K_xBiO_3$[11].

Specific heat, magnetic susceptibility, high-field electrical resistivity, crystallography, and local structural analysis all point to the emergence of a collective electronic state in $KNi_2Se_2$ that likely results from mixed valency and proximity to a localized charge-density-wave (CDW) state. The large low-temperature electronic specific heat coefficient, $\gamma$, indicates significant electronic entropy mass enhancement, but only at low temperatures. The magnetic susceptibility exhibits temperature-independent Pauli paramagnetism with no long-range magnetic order observed by neutron diffraction. Structural studies reveal local CDW/bonding fluctuations at high temperature ($T = 300$ K) that disappear on cooling. Taken together, our results imply the formation of a coherent many-body, heavy-fermion state in $KNi_2Se_2$ in which mixed valence plays a key role.

## Methods

$KNi_2Se_2$ was synthesized from the elements as a polycrystalline, lustrous, purple-red powder, as previously described[12]; the same batch of sample was used for all measurements. The moderately air-sensitive powders required great care in specimen preparation (dry argon or He atmospheres) for further characterization.

Physical properties (magnetic susceptibility, heat capacity, and transport) were measured from 1.8 K to 300 K using a Physical Properties Measurement System (PPMS, Quantum Design, Inc.). The magnetic susceptibility was determined by, $\chi \approx (M_{2\,T} - M_{1\,T})/ 1$ T, in order to remove the contribution from a 0.9 wt% ferromagnetic Ni impurity. A large sample (~100 mg) was measured to minimize small diamagnetic signals from the sample holder. The specific heat was measured using a semi-adiabatic heat-pulse technique in a PPMS equipped with a dilution refrigerator for measurements below $T < 1.8$ K.

Magnetoresistance was measured at the National High Magnetic Field Laboratory (NHMFL, Tallahassee, FL); low frequency AC longitudinal resistance measurements were performed in 4-point geometry in a $^3He/^4He$ superconducting magnet system (SCM-II at the NHMFL) in a perpendicular magnetic field up to $B = 18$ T (ramp speed 0.3 T/min). Samples were mounted and thermally anchored with standard 16 pin DIP sample mounts in an $^3He$ exchange gas atmosphere and cooled to base temperature, $T = 0.3$K within 2 hours. Constant AC current ($f \sim 12$ Hz and $I = 1.2$ mA rms) was applied with a Keithley Instruments, Inc. 2612A source meter and voltages were recorded with Stanford Research Systems, Inc. 830 lock-in amplifiers over shielded twisted pair connections.

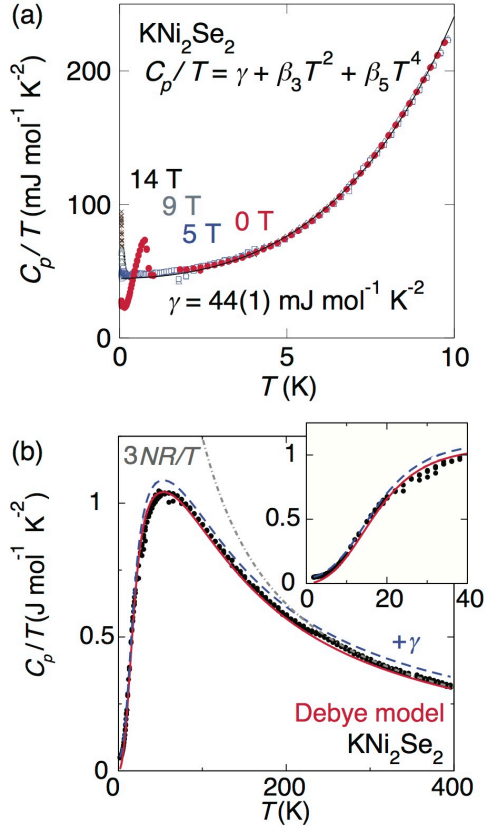

**Figure 1.** (Color online) (a) The low-temperature specific heat of $KNi_2Se_2$ shows a large and field-independent Sommerfeld constant ($\gamma$). (b) The normalized specific heat of $KNi_2Se_2$ (circles), measured from 2 K to 400 K, is well described by a double-Debye model (solid line) at high temperatures without an electronic contribution. While the inclusion of a large electronic term [$\gamma = 44(1)$ mJ mol$^{-1}$ K$^{-2}$] significantly improves the fit at low-temperatures (inset), the high-temperature specific heat is then over-estimated (dashed line). Thus, $\gamma$ is small at high temperatures but large at low temperatures, with the crossover occurring around $T \approx 20$ K (arrow, inset). The Dulong-Petit limit of the lattice contribution ($3NR$) is reached at by 300 K (dot-dashed line).

Time-of-flight spallation neutron scattering experiments were conducted on the HIPD and NPDF instruments at the Los Alamos Neutron Science Center (LANSCE), Los Alamos National Laboratory (LANL). Rietveld analysis of NPD data from NPDF were performed using EXPGUI/GSAS[13,14]. Pair distribution function (PDF) analysis was performed on the total neutron scattering data and $G(r)$ were extracted with $Q_{max} = 35$ Å$^{-1}$ (NPDF) and 30 Å$^{-1}$ (HIPD) using PDFgetN[15]. Least-squares (LS) refinements to the PDF from NPDF were performed using PDFgui[16] after defining instrumental resolution parameters from data collected on polycrystalline Si; Reverse Monte Carlo (RMC) fits to the PDF and Bragg profile from NPDF were performed using RMCProfile[17], with a small penalty imposed for breaking local [NiSe$_4$] tetrahedral coordination. Structures were visualized using VESTA[18]. Electron diffraction results were obtained from a JEOL 2100F transmission electron microscope (200 kV) at Brookhaven National Laboratory equipped with a Gatan, Inc. liquid-helium sample holder.

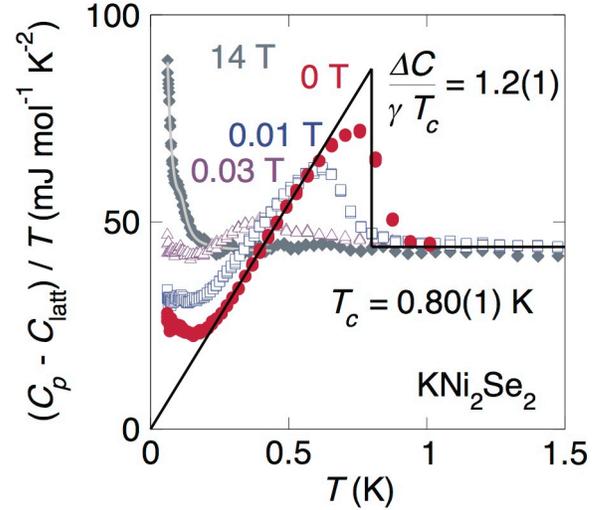

**Figure 2.** (Color online) The electronic specific heat reveals the emergence of bulk superconductivity below $T_c = 0.80(1)$ K under zero-field cooling (dark solid line: equal entropy construction), which is suppressed by a small magnetic field. A high external magnetic field yields a sharp upturn at $T < 0.15$ K described by Schottky anomalies corresponding to impurity (~$10^{-5}$) and nuclear spins (light solid line).

### Results and Discussion

### Physical Properties

$KNi_2Se_2$ crystallizes in the ThCr$_2$Si$_2$ structure-type common to the high-temperature iron-based superconductors[12,19]. The lattice comprises layers of edge-sharing [NiSe$_4$] tetrahedral units, which brings the nickel atoms into close proximity and allows for direct Ni–Ni bonding interactions[12]. The closest Se-Se distances are 3.9 Å, which is too long for a Se$_2^{2-}$ dimer. As such, there is a formal "Ni$^{1.5+}$" mixed valence of the nickel atoms, deduced from K$^+$ and Se$^{2-}$.

The specific heat at $T < 10$ K (**Fig.1a**) reveals a large linear electronic contribution to the low temperature heat capacity [$\gamma = 44(1)$ mJ mol f.u.$^{-1}$ K$^{-2}$]. This value corresponds to a mass enhancement between $m^*/m_b = 18$ (assuming 1.5 carriers/Ni and a spherical Fermi surface) and $m^*/m_b = 6$ (compared to a previous density functional theory calculation)[12]. This mass enhancement is field-independent up to $\mu_0 H \leq 14$ T.

Such mass enhancement is only present at low temperature: the measured specific heat above $T > 300$ K is consistent with the Dulong-Petit limit for phonons ($3NR$) with an electronic contribution corresponding to little or no mass enhancement, as shown in **Fig. 1b**. Inclusion of the low-temperature, mass-enhanced Sommerfeld coefficient greatly over-represents the measured specific heat at 400 K. Thus the specific heat data imply that a significant mass enhancement only sets in at low temperature. To quantify the temperature at which the low-temperature enhancement of the electronic specific heat appears, we estimated the total lattice contribution to the specific heat using a "double-Debye" model.

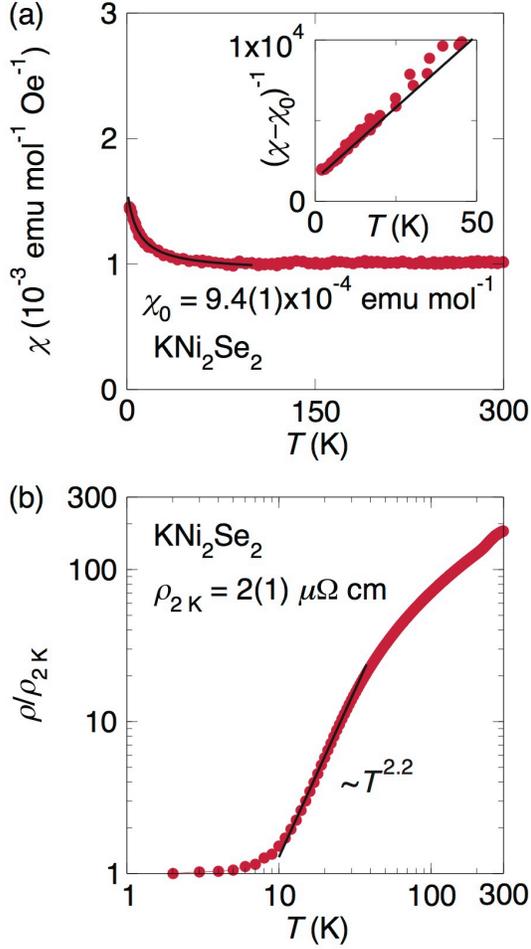

**Figure 3.** (Color online) (a) The magnetic susceptibility indicates temperature independent Pauli paramagnetism [solid line and inset: Curie tail corresponding to 0.5(1) mol% of an $S = 1$ impurity]. (b) Metallic resistivity data reveal a high residual resistivity ratio ($\rho_{300K}/\rho_{2K} \sim 170$) and a nearly quadratic temperature dependence for $T < 40$ K (solid line). The deviation from quadratic behavior below 10 K likely results from scattering off grain boundaries.

The Debye description of the specific heat of a harmonic lattice is given by,

$$C_v = 9RN\left(\frac{T}{\Theta_D}\right)\int_0^{x_D} \frac{x^4 e^x dx}{(e^x - 1)^2}, \quad (1)$$

where $R$ is the gas constant, $N$ is the total number of atoms per formula unit, $\Theta_D$ is the Debye temperature, and $x_D = \Theta_D/T$.[20] For real materials with complex structures and connectivity, this model fails to capture the temperature dependence of the collective lattice degrees-of-freedom. A simple perturbation of this model is to assume two independent Debye lattices: a "double-Debye" model. This is given by the expression,

$$C_v = \left[9R(N-s)\left(\frac{T}{\Theta_{D1}}\right)\int_0^{x_{D1}} \frac{x^4 e^x dx}{(e^x - 1)^2}\right] + \left[9Rs\left(\frac{T}{\Theta_{D2}}\right)\int_0^{x_{D2}} \frac{x^4 e^x dx}{(e^x - 1)^2}\right], \quad (2)$$

where each sublattice has a unique Debye temperature ($\Theta_{D1}$, $\Theta_{D2}$, $x_{D1}$, $x_{D2}$) and a factor to partition the number of participating modes into each sublattice ($s$ is the number of modes in the second sublattice). This expression provides an excellent fit to the high-temperature specific heat of $KNi_2Se_2$, illustrated in **Fig.1(b)**.

By fitting the specific heat in the high-temperature region to reduce the relative contribution of the electronic contributions (fit as $C_p/T$), we extracted the Debye temperatures by assuming 3 oscillators in one lattice, and 2 in the other ($N = 5$, $s = 2$). The values are summarized in **Table I** for fits over different temperature ranges. Fitting the specific heat above $T > 30$ K, we observe a finite electronic contribution [$\gamma \sim 14(1)$ mJ mol$^{-1}$ K$^{-2}$] that is comparable to other structurally analogous nickel compounds [e.g., LaNiPO, $\gamma_{0T} = 5.2(1)$ mJ mol$^{-1}$ K$^{-2}$][21]; however, this value is significantly reduced from the value determined for $KNi_2Se_2$ at low temperature [$\gamma = 44(1)$ mJ mol$^{-1}$ K$^{-2}$]. A crossover appears at low temperature [$T \approx 20$ K, inset **Fig.1(b)**] where inclusion of the large linear $\gamma$ contribution describes extra electronic entropy not captured by the lattice. Above $T \approx 20$ K, inclusion of the $\gamma = 44(1)$ mJ mol$^{-1}$ K$^{-2}$ electronic term overestimates the total specific heat, illustrating that the mass enhancement only occurs at low temperatures.

Superconductivity emerges from this heavy-mass electronic state; zero-field cooling reveals a clear $\lambda$-anomaly in the specific heat at $T_c = 0.80(1)$ K, indicative of bulk superconductivity (**Fig.2**). The application of a large external magnetic field ($\mu_0 H = 14$ T) suppresses the superconducting transition and produces an upturn in the specific heat as $T < 0.15$ K. This entropy is well described by Schottky anomalies corresponding to $\sim 10^{-5}$ impurity spins (i.e. a paramagnetic Curie tail) and nuclear spins ($C_p = c_{nuc}T^{-2}$). From the temperature dependence of the suppression of the superconducting transition under small magnetic fields, we estimate an average $H_{c2}(0K) \approx 0.05(1)$ T from the two-fluid model, $H_{c2}(T) = H_{c2}(0K)[1 - (T/T_c)^2]$. From this upper critical field, the average zero-temperature coherence length of the superconducting state is determined to be $\xi(0) = [\phi_0/(2\pi H_{c2})]^{1/2} = 80(1)$ nm. By assuming a spherical Fermi surface, we can estimate an effective band mass from the superconducting transition temperature, $T_c$ and coherence length, $\xi$. First, the Fermi velocity can be calculated from $T_c$ and $\xi$ using the relation $v_F = \frac{k_B T_c \xi}{0.18 \hbar} = 46500 \frac{m}{s}$. Then, the Fermi wavevector can be estimated from the carrier density, $n =$ carriers per unit cell volume, using $k_F = (3n\pi^2)^{1/3}$. If one assumes 1.5 carriers per Ni, then $k_F = 0.96(1)$ Å$^{-1}$, while if one instead assumes all 33 valence electrons per formula unit contribute, then $k_F = 2.1(1)$ Å$^{-1}$. The resulting mass enhancement is then calculated from the relation $m^*_{eff} = \hbar k_F / v_F$, and ranges from $m^*_{H_{c2}} = 24$ to $50$ $m_e$. This range indicates significant mass enhancement and is in reasonable agreement with the value of $m^* = 6$ to $18 m_e$, calculated from the Sommerfield coefficient, especially considering the limitations of a spherical Fermi surface approximation in this layered material.

Further, only ~50% of the electrons enter the superconducting state ($\gamma_{resid} / \gamma$); the origin of the residual electronic entropy, $\gamma_{resid} = 22(3)$ mJ mol f.u.$^{-1}$ K$^{-2}$, below $T_c$ remains unclear, but it may be attributed either to impurity states in a superconductor with nodes (*e.g.* cuprates)[22], hinted at given the nearly quadratic dependence of the specific heat for $T < T_c$, or to microscopic inhomogeneity[5]. In either case, the specific heat jump at $\Delta(C - C_{latt})/(\gamma T_c) = 1.2(1)$ is proof of bulk superconductivity arising from a heavy-electron-mass state.

The magnetic susceptibility of KNi$_2$Se$_2$ [**Fig.3(a)**] is nearly temperature independent, consistent with previous reports of Pauli paramagnetism[12,19]. The temperature-independent susceptibility [$\chi_0 = 9.4(1)\ 10^{-4}$ emu mol f.u.$^{-1}$ Oe$^{-1}$] is indicative of a significant density of states at the Fermi level [$\chi_{Pauli} = \mu_B^2\ g(E_F)$], with no evidence for Curie-Weiss behavior that would be expected from the presence of localized magnetic states. The susceptibility does exhibit a weak upturn below $T < 50$ K, well described by a Curie tail [inset **Fig.3(a)**] corresponding to 0.5(1) mol% of a $S = 1$ impurity (*e.g.*, Ni$^{2+}$).

The monotonic decrease of the electrical resistivity on cooling indicates metallic transport [**Fig.3(b)**]. The residual-resistivity ratio [$\rho_{300K}/\rho_{2K} \sim 175$, $\rho_{2K} \sim 2(1)$ $\mu\Omega\cdot$cm] is unusually high for a measurement on a low-density polycrystalline pellet, and is instead comparable to high-purity copper ($\rho_{300K}/\rho_{2K} \sim 10^2$-$10^5$)[23] or to single-crystalline RuO$_2$ ($\rho_{300K}/\rho_{2K} \sim 10^2$ - $10^3$)[24]. Furthermore, the resistivity is approximately quadratic for $T \lesssim 40$ K [$\rho(T) \sim T^{2.2}$]. Below 10 K, the resistivity is dominated by grain boundary and other residual contributions, making it difficult to assess the intrinsic behavior in that region. Future work is needed to further reduce impurity scattering and determine whether this $T^2$ behavior continues down to the lowest temperatures. The Kadowaki-Woods ratio relates temperature coefficient in such a quadratic regime ($A$) to the electronic specific heat, $A/\gamma^2$, and is typically ~$10^{-5}$ $\mu\Omega$·cm [mol K$^2$ mJ]$^2$ for heavy-fermion systems[25,26]. For KNi$_2$Se$_2$, we calculate $A/\gamma^2 \sim 1.2\times10^{-5}$ $\mu\Omega$·cm [mol K$^2$ mJ]$^2$ for $T < 40$ K. This scaling relation aligns the metallic properties of KNi$_2$Se$_2$ with heavy-fermion behavior.

Magnetoresistance (MR) measurements do not show any sign of magnetically induced transitions (beyond the suppression of superconductivity) up to $\mu_0 H = 18$ T [**Fig.4(a)**]. Instead, the changes in curvature of the MR are well described by a nearly-compensated two-band metal which predicts $\rho/\rho_0 = 1+[(1 + bB^2) / (1 + c\ b^2 B^2)]$[27]. In the Drude form, the carrier mobility is expressed as $\mu_{Drude} = q_i \tau_i / m_i$, where $q_i$, $\tau_i$, and $m_i$ are the carrier charge, scattering time, and effective mass, respectively. For free electrons in a two-channel metal, the magnetoresistance, $\rho_{xx}(B)$ is given by,

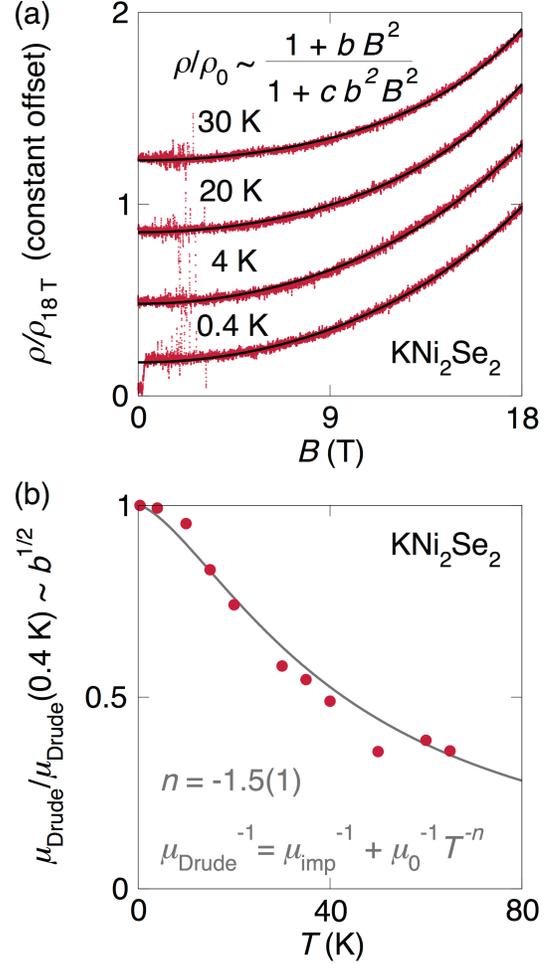

**Figure 4.** (Color online) (a) The high-field magnetoresistance ($\rho/\rho_{18T}$) shows a change in curvature as a function of temperature (black lines: fit to the two-carrier Drude model, **Eqn.7**). (b) Temperature dependence of the carrier mobility extracted from the high-field magnetoresistance reveals the formation of a coherent heavy-band below $T \approx 40$ K. The solid line reflects a fit to Matthiessen's rule for carrier mobilities (**Eqn. 8**).

$$\rho_{xx} = \rho = K \cdot B \cdot \frac{\mu_1 B + n\mu_2 B + \mu_1\mu_2 B^2(\mu_2 B + n\mu_1 B)}{(\mu_1 B + n\mu_2 B)^2 + (1-n)^2 \mu_1^2 \mu_2^2 B^4}, \quad (3)$$

which, by combining terms reduces to,

$$\rho = K \cdot \frac{\mu_1 + n\mu_2 + \mu_1\mu_2 B^2(\mu_2 + n\mu_1)}{(\mu_1 + n\mu_2)^2 + (1-n)^2 \mu_1^2 \mu_2^2 B^2} \quad (4)$$

where $\mu_1$ and $\mu_2$ are the magnitude of the carrier mobilities for each channel, $B$ is the applied magnetic field, $K$ is a constant, and $n$ is the carrier ratio ($n_1/n_2$). In a compensated metal, $n = 1$. In general, without approximation, at $B = 0$, $\rho_0 = K \cdot (\mu_1 + n\mu_2)^{-1}$. Hence, the magnetoresistance ratio is given by [when dividing a simplified Eqn.(4) by $(\mu_1 + n\mu_2)^2$],

$$\frac{\rho}{\rho_0} = 1 + \frac{n(\mu_1 + \mu_2)^2 \mu_1 \mu_2 B^2}{(\mu_1 + n\mu_2)^2 + (1-n)^2 \mu_1^2 \mu_2^2 B^2}, \quad (5)$$

without approximation. Taking

$$b = \frac{n(\mu_1 + \mu_2)^2}{(\mu_1 + n\mu_2)^2} \mu_1 \mu_2 \quad \text{and} \quad c = \frac{(1-n)^2 (\mu_1 + n\mu_2)^2}{n^2 (\mu_1 + \mu_2)^4} \quad \text{gives,}$$

$$\frac{\rho}{\rho_0} = 1 + \frac{bB^2}{1 + cb^2 B^2}. \quad (6)$$

Thus we fit the magnetoresistance data to the expression,

$$\frac{\rho}{\rho_{18T}} = \text{OMR}\left(1 + \frac{bB^2}{1 + cb^2 B^2}\right), \quad (7)$$

where OMR is a scale factor to account for the fact that the magnetoresistance data was scaled to the high field value. This equation provided an excellent fit to the data at all temperatures [**Fig.4(a)**]. The curvature of the data required that values of $c$ take on negative values, which may be physically unrealistic since $c$, as defined above, must be a positive constant. This inconsistency has been observed in other materials, such as polycrystalline copper, and has been attributed to the existence of sharp edges on the Fermi surface.[27,28] The temperature dependence of the magnetoresistance indicates a significant change in the electronic state on cooling below $T \lesssim 40$ K. This change visibly manifests as a change of curvature of $\rho/\rho_{18\,T}$ with temperature, as reflected in the trends observed in $\mu_{\text{Drude}}$ (defined as $b^{1/2}$), $c$, and OMR (**Table II**).

The most relevant parameter is the relative carrier mobility, $\mu_{\text{Drude}}$, which exhibits a large increase on cooling below $T \approx 40$ K [**Fig.4(b)**]. The temperature dependence is reasonably well described by Matthiessen's rule for the mobility, if one assumes that there are contributions from an impurity limiting mobility ($\mu_{\text{imp}}$) and an intrinsic temperature dependent mobility ($\mu_0 T^{\,n}$) of the coherent electronic state,

$$\frac{1}{\mu_{\text{Drude}}} \approx \frac{1}{\mu_{\text{imp}}} + \frac{1}{\mu_0 T^{\,n}} \quad (8)$$

where $n = -1.5(1)$, $\mu_{\text{imp}} = 1100$ cm$^2$ V$^{-1}$ s$^{-1}$, and $\mu_0 = 3(1) \times 10^6$ cm$^2$ V$^{-1}$ s$^{-1}$. The temperature dependence ($n$), resembles that of a high-mobility semiconductor.[29] Since the carrier mobility is proportional to the mean scattering time divided by the effective band mass, the large increase in effective mass at low temperatures must be compensated by an even larger increase in mean scattering time.

**Structure analysis**
Structural analysis of KNi$_2$Se$_2$ provides direct evidence for fluctuating CDWs at high-temperature that disappear on cooling. High resolution, and high momentum transfer ($Q$) time-of-flight neutron powder diffraction (NPD) data and Rietveld analyses show no change in the $I4/mmm$ symmetry from 300 K to 15 K or evidence for long-range magnetic order (**Fig.5**, **Table III**). Neutron pair distribution function analysis (PDF) from the HIPD instrument shows a spatially incoherent lattice distortion at 300 K, corresponding to multiple Ni–Ni distances between 2.5 Å < $r$ < 3.0 Å (**Fig.6**). Curiously, the PDF signature of multiple Ni–Ni distances gradually disappears on cooling; the peak positions do not change, but rather, the populations decrease, with full loss of the local CDW occurring between $T = 25$ K and $T = 15$ K (**Fig.6**). The disappearance of a structural distortion on *cooling* is highly unusual, and to our knowledge has only been previously observed in a few other systems: the high $zT$ thermoelectric lead chalcogenides[30] which exhibit giant anharmonic phonon scattering[31] resulting from stereochemical activity of the Pb$^{2+}$ lone-pair[32], the perovskite manganites which form an insulating charge-ordered (disproportionated Mn$^{3+}$/Mn$^{4+}$) state at high temperature that evolves into a ferromagnetic metal on cooling[33,34], and PbRuO$_3$ which undergoes an untilting of RuO$_6$ octahedra when orbital order sets in[35].

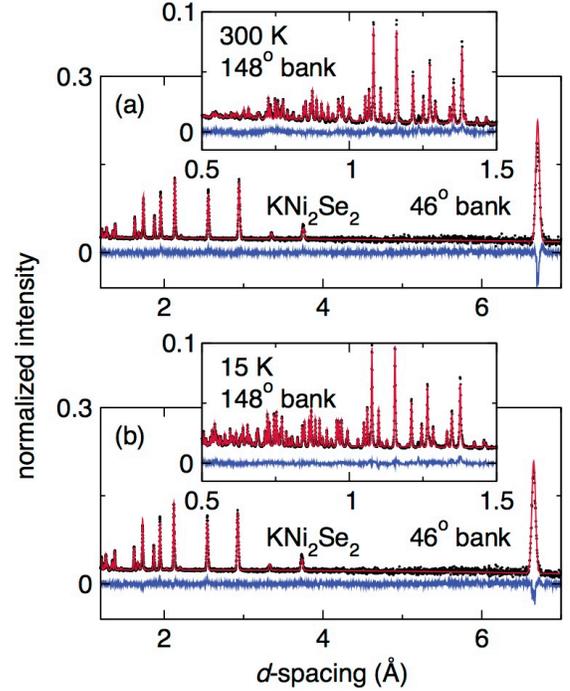

**Figure 5.** (Color online) Time-of-flight neutron diffraction of KNi$_2$Se$_2$ at (a) $T = 300$ K and (b) 15 K from forward-scattering and back-scattering (inset) detector banks. Data (dots) and Rietveld analyses (lines) show no deviations (difference, offset line) from the ideal $I4/mmm$-symmetry of the ThCr$_2$Si$_2$-type structure at both temperatures.

Calculation of various partial pair-wise correlations from the crystallographic structure indicates that the largest discrepancy arises from the Ni–Ni distances; least-squares fitting of the PDF illustrates that the $I4/mmm$ unit-cell cannot capture the local arrangement of atoms in KNi$_2$Se$_2$ at $T = 300$ K (**Fig.7**). Various locally ordered structural models, including distortions of the maximal subgroups of the $I4/mmm$ unit-cell, also fail to describe the PDF at 300 K.

To characterize the splitting of Ni–Ni distances at $T$ = 300 K, we employed large-box (24000 atoms, ~$5\times10^5$ Å$^3$) Reverse Monte Carlo (RMC) simulations of the *total* scattering data from the NPDF instrument by simultaneously fitting the starting configuration, based on the *I4/mmm* unit-cell, to both the Bragg profile and the PDFs [**Fig.8(a)**]. The RMC simulation of the $T$ = 15 K data yields an equivalent fit to the PDF as least-squares refinement of the average crystal structure [**Fig.8(b)**]. From the RMC analysis, the symmetric distribution of Ni–Ni distances at $T$ = 300 K reveals three distinct populations of Ni–Ni bonds [**Fig.8(a)**, inset], in contrast to the single Ni–Ni bond length at $T$ = 15 K. As a control, Bragg profiles and PDFs, were simulated from the average crystal structure with anisotropic Debye-Waller factors and the appropriate instrumental parameters (*e.g.*, $Q_{max}$); these simulated scattering profiles were then fit using the same RMC methods to provide a direct comparison and plotted in **Fig.8(a)** and **(b)** to further illustrate how the local distortions at $T$ = 300 K cannot be described by anisotropic harmonic atomic displacements.

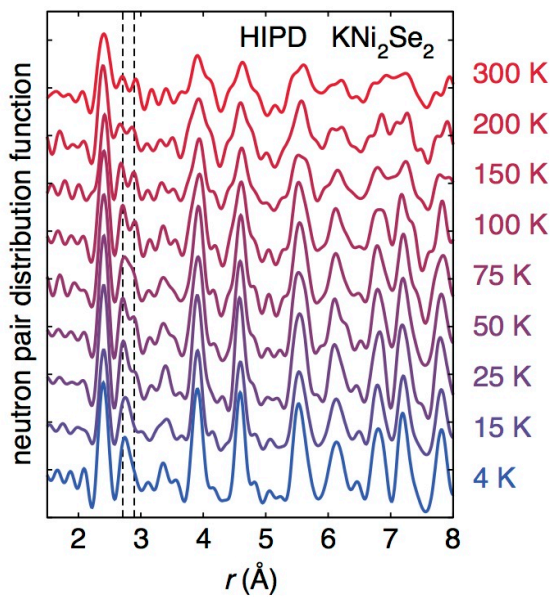

**Figure 6.** (Color online) Temperature dependent pair distribution function analysis of KNi$_2$Se$_2$ reveals multiple Ni–Ni distances at $T$ = 300 K (dashed lines at $r$ ~ 2.7 Å and 2.9 Å), yet only one Ni–Ni distance at $T$ = 4 K. The loss of multiple Ni-Ni distances occurs between $T$ = 25 K and $T$ = 15 K.

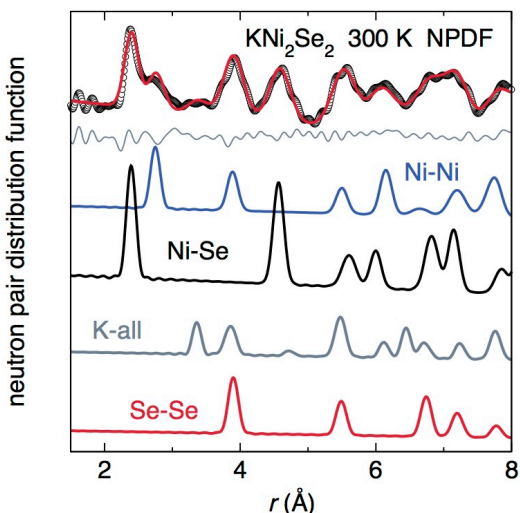

**Figure 7.** (Color online) Least-squares fitting of the *I4/mmm* crystal structure of KNi$_2$Se$_2$ to the PDF (solid line), with partial pair-wise contributions below, does not provide an adequate description of the Ni–Ni or Ni–Se distances at $T$ = 300 K.

From the RMC analysis, the Ni–Ni displacements appear to be spatially incoherent. A glimpse of one [Ni$_2$Se$_2$] layer within the RMC supercell shows a random distribution of long and short Ni–Ni bonds [**Fig.8(c)**]. While a the short Ni–Ni distance requires the Ni atoms to move within the *ab* plane [**Fig.8(d)**], a long Ni–Ni distance is generated by displacement of the Ni atoms along *c* [**Fig.8(d)**]. Folding the 24000 atom configuration back into the crystallographic unit-cell generates the atomic "point-clouds" represented in **Fig.9(a)**. On average, these point-clouds qualitatively resemble anisotropic Debye-Waller factors (**Table III**), as previously reported.[12] From the RMC simulation of the experimental data collected at $T$ = 300 K, the histograms of Ni atom displacements away from the ideal crystallographic site (along *c* or along *a*) [**Fig.9(b)**] are described by a normal distribution, with crystalline anisotropy such that $\Delta z > \Delta x$. At $T$ = 15 K, the Ni displacements are nearly isotropic ($\Delta z \approx \Delta x$).

Additionally, the bond angles are consistent with a lack of ordered pattern formation. The normal distribution of tetrahedral angles ($\theta$) with Ni–Ni distance further suggests that the origin of the structural distortion is not a single-ion effect [**Fig.10(a)**]. If a Jahn-Teller distortion were responsible for the distribution of Ni–Ni distances, the configuration would reveal distinct populations of Se–Ni–Se angles,[36] especially when plotted as a function of Ni–Ni distance. There is a correlation between the Ni–Se–Ni angle ($\phi$) and the Ni–Ni distance [**Fig.10(b)**], as expected from geometric considerations: shorter bonds decrease $\phi$. Taken together, these analyses illustrate $T$ = 300 K is locally disordered; however, the structure at $T$ = 15 K is well described by the average crystal structure.

Electron diffraction directly shows the absence of local pattern formation within the *ab* plane: no additional reflections are observed from $T$ = 11 K to 300 K in selected area electron diffraction patterns of in KNi$_2$Se$_2$, in either the zero order Laue zone (ZOLZ) or first-order Laue zone (FOLZ) when viewed along the ⟨001⟩ zone axis (**Fig.11**). While the pair distribution function is an ensemble structure probe, the SAED is sensitive to local ordering of atom positions from the appearance or absence of distinct symmetry elements that add, remove, or move the observed reflections. Using a ~100 nm spot size, SAED patterns collected on the same region of the sample (**Fig.11**) do not show any evidence for symmetry lowering reflections or enlargement of the unit-cell that would be indicative of local ordering of such distortions at any of the temperature.

Sensitivity of the sample to both the incident beam and sample processing prevented a reduction of the illuminated region and reorientation along other zone axes. These data imply that the Ni–Ni displacements are dynamic and/or entirely uncorrelated between unit-cells within the *ab* plane [**Fig. 8(c)**], in contrast to the coherent CDWs observed in structurally-related compounds such as $KCu_2Se_2$ and $SrRh_2As_2$[37,38].

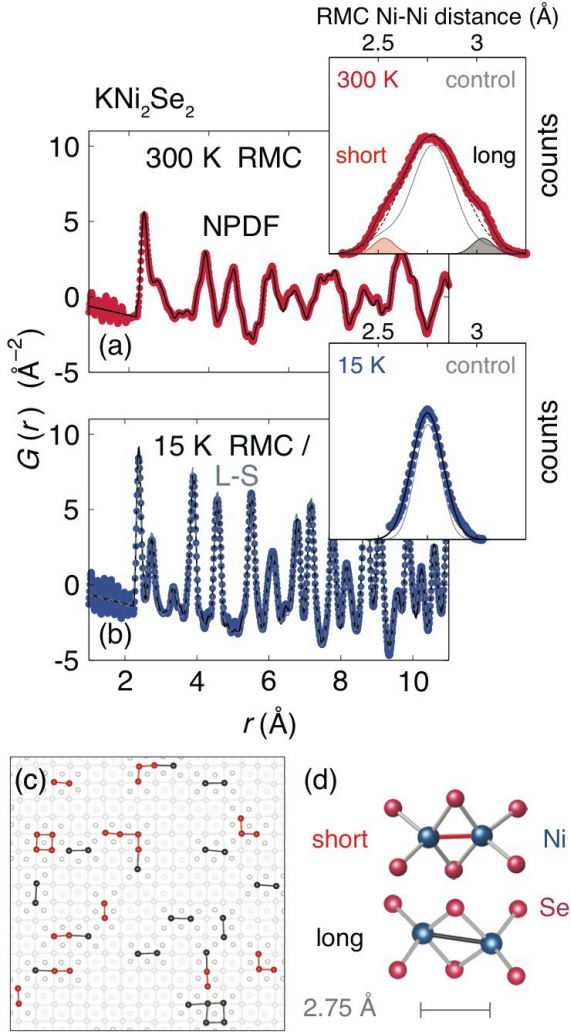

Extraction of the bond valence sums[39] from Ni-Se distances within [$NiSe_4$] tetrahedra reveals a single population of Ni species, illustrated in **Fig.12(a)**. The single population, centered around BVS = 1.9(3), is comparable to the control simulation (average) with a BVS = 2.0(2), rather than distinct populations from charge disproportionation, as seen in $CuMn_2O_4$,[36] or the value expected from a formal

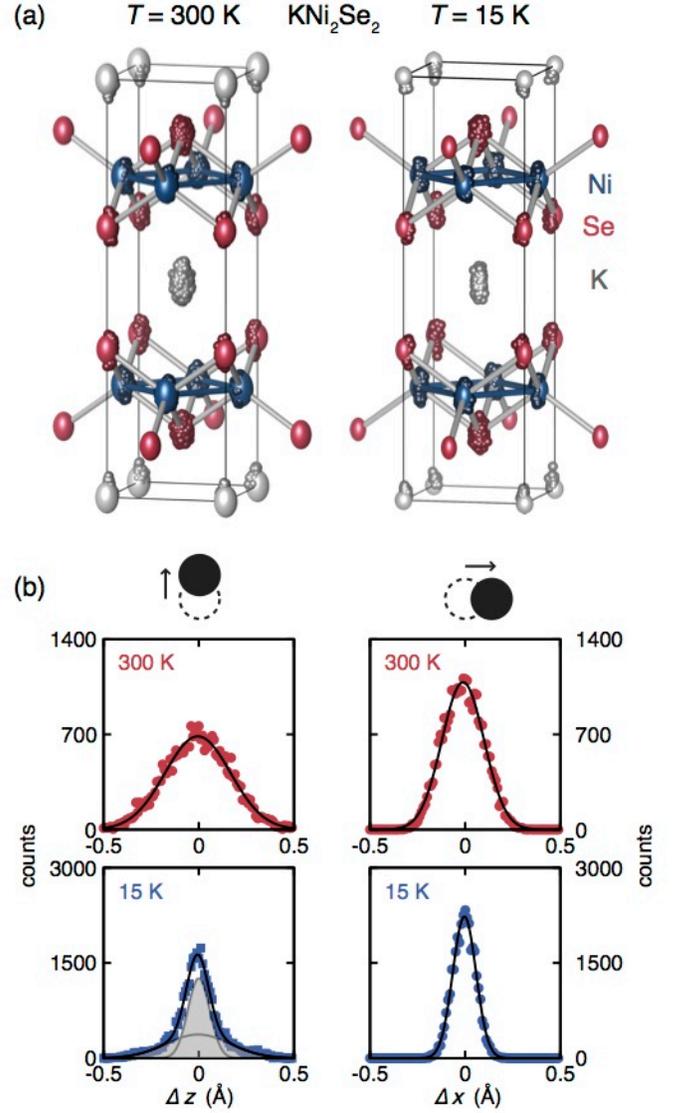

**Figure 8.** (Color online) Fits of the PDF (symbols: data, dark line: fit) at (a) $T$ = 300 K and (b) $T$ = 15 K by RMC simulation (solid line) or by least-squares refinement of the average *I4/mmm* crystal structure (dashed grey line, $T$ = 15 K only). Insets: the RMC Ni–Ni distance distributions at $T$ = 300 K shows three types of bond lengths, in contrast to the single distribution from a control RMC simulation of a PDF calculated from the average crystal structure (solid grey line). A single bond length population is found at $T$ = 15 K. (c) A [$Ni_2Se_2$] layer extracted from the RMC supercell shows no locally ordered patterns of short ($r$ < 2.48 Å, red) and long Ni–Ni bonds ($r$ > 3.1 Å, black). (d) Representative short and long Ni–Ni bonds illustrate the concomitant Se displacements for both distortions (represented to scale).

**Figure 9**. (Color online) (a) Atomic "point clouds" (small spheres), generated by folding the 24000 atom configuration back into the *I4/mmm* unit-cell, resemble anisotropic Debye-Waller factors (large thermal ellipsoids at 99% probability) on *average*, consistent with the Rietveld analysis (**Table III**) and the previously reported crystal structure.[13] (b) Displacements of the Ni position from the crystallographic position along the $z$ axis (perpendicular to the [$Ni_2Se_2$] layers, left) and along $x$ (in plane, right) indicate that the Ni positions are significantly more displaced along $z$ at $T$ = 300 K than along $x$. The atomistic representation at $T$ = 15 K more closely represents an isotropic atomic displacement parameter, as the majority of Ni atoms (~55%) exhibit $\Delta z \approx \Delta x$; a second, broad normal distribution describes the remaining ~45% of Ni atoms with larger $z$ displacements (histogram from RMC: symbols, normal distributions: lines).

$Ni^{1.5+}$ valence. Therefore, we infer that the extra $0.5e^-$ / Ni is delocalized in band. This suggests that the origin of the structural distortions is not a single-ion effect (*e.g.*, a Jahn-Teller distortion), which would be reflected as multiple populations in a BVS. Since the BVS does not take into consideration Ni–Ni bonds, we also analyzed the [NiSe$_4$] tetrahedral volumes. Compared to a harmonic distribution of atom positions generated from the control RMC simulation, the $T = 300$ K configuration displays a second normal distribution tetrahedral volumes [**Fig.12(b)**]. The expanded volume of the second population is consistent with the presence of additional (repulsive) negative charge, and resembles the local distortions depicted in **Fig.8(a)** and **(b)**. Whereas, at $T = 15$ K, a single population of tetrahedral volumes is recovered.

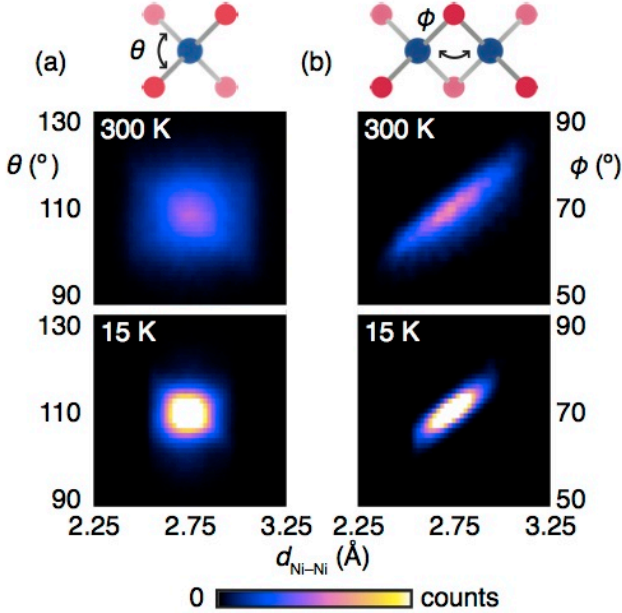

**Figure 10.** (Color online) (a) The two-dimensional histograms of tetrahedral Se-Ni-Se angles ($\theta$, left, scale: 120 counts) show that the tetrahedral angles do not anomalously correlate with Ni–Ni bond length or with temperature. (b) The two-dimensional histograms of Ni-Se-Ni angles ($\phi$, right, scale: 60 counts) show that the out-of-plane Ni-Se-Ni angle increases with Ni–Ni distance, as geometrically expected.

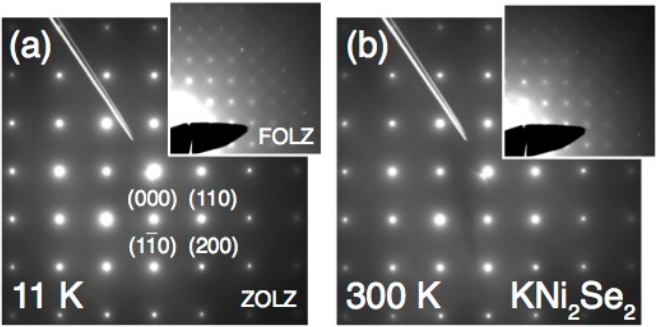

**Figure 11.** (Color online) Representative SAED patterns from the same area of the sample reveal an absence of local distortions orthogonal to the $\langle 001 \rangle$ zone axis at (a) $T = 11$ K and (b) $T = 300$ K.

**Relation to traditional heavy-fermion behavior**

Specific heat and pair-distribution function analysis indicate that there is a large increase in the effective mass of conduction electrons at the same time the CDW disappears, below $T \approx 20$ K, suggesting that the increase in effective mass and loss of a CDW are interrelated. Further, there are large changes in resistivity and carrier mobility below $T \approx 40$ K that suggest the emergence of a coherent many-body electronic state. The most exciting possibility to explain these observations is the formation of a heavy-fermion state that is driven by hybridization of localized charges with conduction electrons. In traditional heavy-fermion compounds, there are localized electronic states that arise from narrow bands derived from $4/5f$ or $3d$ orbitals[4,5,40]. Strong on-site electrostatic repulsion (in the form of a Hubbard $U$) normally drives such systems to magnetism [**Fig.13(a)**, *e.g.*, ferromagnetism or a spin-density wave (SDW)]; however, in the case of heavy-fermion behavior[5], Kondo screening and hybridization between localized magnetic and delocalized conduction band levels instead result in the formation of a non-magnetic ($S = 0$) many-electron state [**Fig.13(a)**, right panel][41,42]. The effective electronic mass enhancement in these compounds is conventionally found to be in excess of $m^*/m > 100$, rather than the factor of 6-18 found for $KNi_2Se_2$.[5]

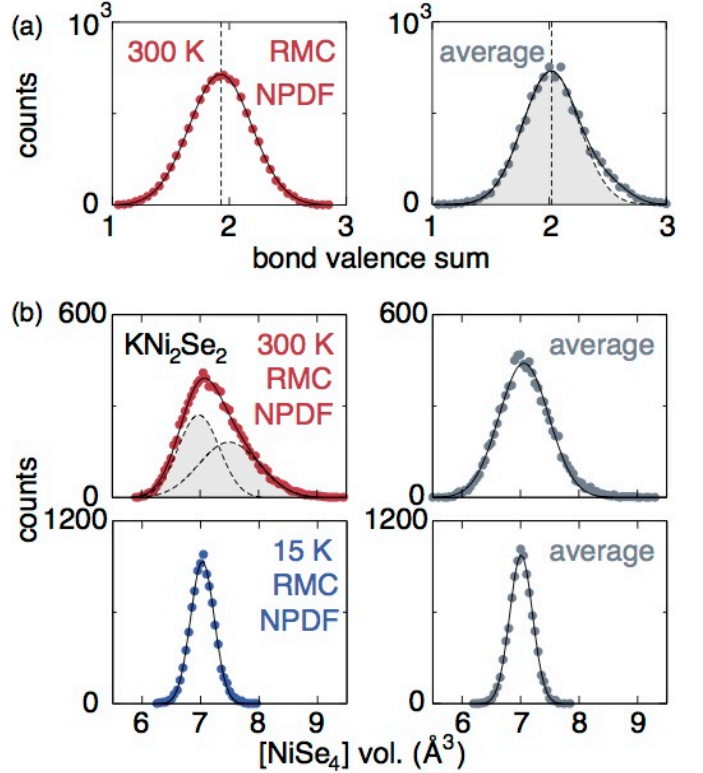

**Figure 12.** (Color online) (a) The histogram of bond valence sums (BVS) calculated from Ni–Se distances of [NiSe$_4$] tetrahedra in the 24000 atom supercell reproduces a single BVS population at $T = 300$ K [black line, centered at 1.9(3) marked by the dashed line, left]. As a control, the same initial supercell was fit to a calculated PDF using the anisotropic Debye-Waller factors obtained from Rietveld refinement (right, labeled "average"). The average BVS reproduces a single population centered at 2.0(2) (dashed line). (b) Histograms of the [NiSe$_4$] tetrahedral volumes extracted from the RMC simulation of the $T = 300$ K data (top, left) reveal two populations at $T = 300$ K (dashed black lines). The control simulation of average structure has only one population by RMC (top, right). The RMC simulation of the $T = 15$ K data (bottom, left) and control simulation of the average structure (bottom, right) have only one population of tetrahedral volumes.

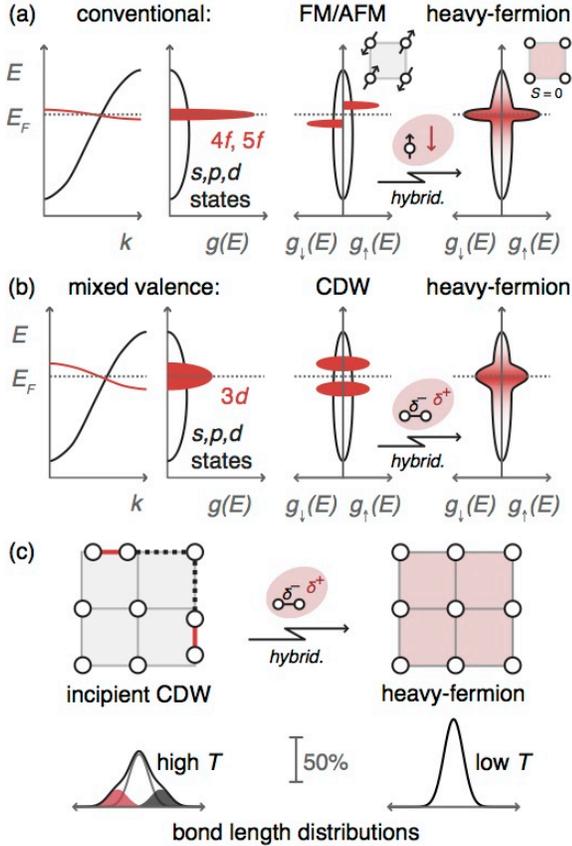

Figure 13. (Color online) (a) Conventional heavy-fermion behavior associated with localized $4/5f$ states are electronically unstable towards magnetism; hybridization (partial red shading) with other states yields a non-magnetic ($S = 0$) many-body state. (b) A CDW instability arises from a half-filled band with appreciable bandwidth compared to the electron-electron interaction strength; hybridization yields a non-CDW many-body state. (c) Incipient CDWs manifest as a *local* modulation of the bond length distribution. Hybridization with conduction electrons (partial red shading) suppresses the CDWs and forms a coherent heavy-mass state with a uniform bond distribution.

In principle, Coulomb interactions between localized, but non-magnetic, states and conduction electrons can also lead to electron mass-enhancement [**Fig.13(b)**]. In the case of a partially-filled and non-magnetic band, charge-localization occurs when the band is electronically unstable against a charge-density wave (CDW)[43]. In a manner analogous to that of nearly magnetic heavy-fermion behavior, screening and hybridization with the conduction electrons can form a non-CDW many-body state. In such a material, there would be a distribution of distinct bond length populations at high temperature, represented schematically in **Fig.13(c),** from incipient CDW formation. As thermal energy decreases, these charge fluctuations disappear and yield a charge-driven heavy-fermion state. Our analysis of the physical properties and structural behavior of $KNi_2Se_2$ suggest the formation of such a charge-mediated, rather than spin-mediated, coherent heavy-fermion state at low temperature, summarized in the phase diagram illustrated in **Fig.14**. In this model, the apparent difference in onset temperature between bulk (specific heat and PDF, $T \approx 20$ K), and transport (resistivity and mobility, $T \approx 40$ K) measurements we observe is then likely due to the continuous phase transition between normal mass (and charge fluctuating), and heavy mass (and no CDW) states: the transport measurements are sensitive to the presence of even small regions of the "heavy mass" state due to the current following the path of least resistance, whereas the bulk probes show a transition only when a majority of the system has collapsed into the heavy mass state. This model also explains the modest 6-18-fold increase over the bare band mass in $KNi_2Se_2$: the band that is electronically unstable toward a CDW has a larger bandwidth than the equivalent bands (unstable toward magnetism) in a typical heavy-Fermion, implying less of a localization effect.

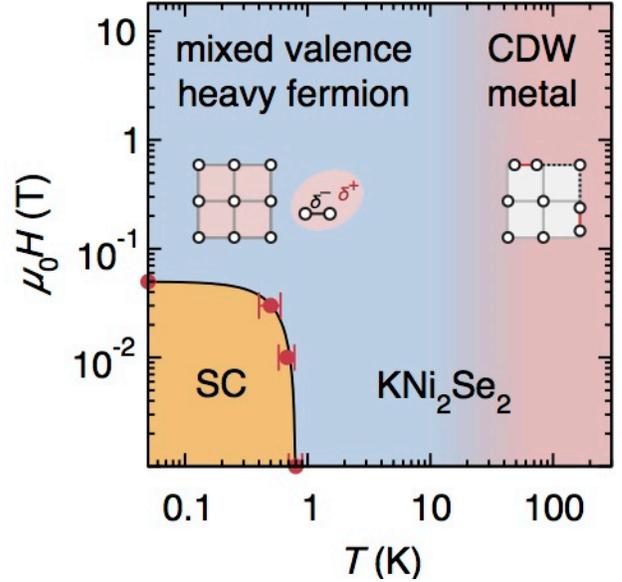

**Figure 14.** (Color online) The temperature/field phase diagram shows the superconducting, field-independent mixed valence heavy-fermion, and incipient charge-density wave metal states of $KNi_2Se_2$.

Many $4/5f$ heavy-fermion materials, as well as the $3d$ heavy-fermion compound, $LiV_2O_4$, exhibit Curie-Weiss-like behavior from localized magnetic states at high temperatures[5,40]; no such behavior, nor magnetic order in NPD data, is found for $KNi_2Se_2$. The field-independence of the electronic contribution to the specific heat in $KNi_2Se_2$ is distinct from the field-dependence exhibited $UPt_3$,[44,44] $CeRu_2Si_2$,[45] $CeCoIn_5$,[46] and $LiV_2O_4$,[40] and further suggests that the mass enhancement in $KNi_2Se_2$ is not directly tied to magnetism. In the filled skutterudite heavy fermion superconductor, $PrOs_4Sb_{12}$, mass enhancement is not directly related dipolar magnetic order, but instead to quadropolar interactions of the localized $4f$ states.[47] In that case, application of a 5 T magnetic field to $PrOs_4Sb_{12}$ is sufficient to induce a transition to antiferro order of quadropolar moments, which indicates a close proximity to magnetism. We observe no such magnetic-field-induced transition in $KNi_2Se_2$ up to $\mu_0H = 18$ T.[48] The mass enhancement is not likely due to exchange-enhancement (*cf.* Pd or TiBe$_2$[20]) either: in the case of Kondo screening, the Wilson ratio [$R_w = (\pi^2 k_B^2 \chi_0)/(3\mu_B^2 \gamma)$] lies near $R_w = 2$,[42] whereas for free electrons, $R_w \sim 1$ and for exchange-enhanced compounds $R_w >> 1$, as elaborated below.

A useful probe of such electron effective-mass enhancement ($m^*/m$) is the Sommerfield constant, $\gamma$, the electronic contribution to the specific heat, since $m^*/m \sim \gamma^*/\gamma$. For a free-electron gas, the Sommerfield $\gamma$ is a metric of the total density of states at the Fermi energy, $\gamma_{free} = \pi^2 k_B^2 g(E_F)/3$. A second probe of the densities of states for a free-electron gas is the Pauli paramagnetic susceptibility, $\chi_{Pauli,free} = \mu_B^2 g(E_F)$. When these two qualities are related through the densities of states, one can write the relation:

$$R_w = \frac{\pi^2 k_B^2}{3\mu_B^2} \frac{\chi_{Pauli,free}}{\gamma_{free}} = 1. \quad (9)$$

Measuring $\chi_{Pauli,free}$ is non-trivial, so the temperature-independent susceptibility is used to compute $R_w$,

$$R_w = \frac{\pi^2 k_B^2}{3\mu_B^2} \frac{\chi_0}{\gamma}. \quad (10)$$

The total temperature-independent susceptibility comprises contributions:

$$\chi_0 = \chi_{Pauli} + \chi_{Landau} + \chi_{orbital} + \chi_{core}. \quad (11)$$

The orbital contribution ($\chi_{orbital}$), is temperature independent, but direct extraction or calculation is highly non-trivial[49]. The core diamagnetic contribution can be easily estimated or calculated ($\chi_{core}$), but is generally small[50]. We therefore take the temperature independent susceptibility, $\chi_0$, as the sum of the conduction electron paramagnetism ($\chi_{Pauli}$) and diamagnetism ($\chi_{Landau}$). The Landau diagmagnetic contribution is $\chi_{Landau} = -\chi_{Pauli} (m/m^*)^2/3$. Using **Eqn. 10**, in the limit that $m^*/m = 1$, $R_w = 2/3$. In the limit of an infinite effective mass ($m^*/m = \infty$), $R_w = 1$.

To achieve a Wilson ratio $R_w > 1$ requires physics beyond the trivial mass-enhancement of free-electron gas, such as Kondo screening in the case of traditional 4/5*f* heavy-fermion compounds. When $R_w > 1$, it qualitatively indicates that the magnetic susceptibility is higher than expected from mass-enhancement. Wilson showed that in the case of Kondo screening, the ratio of the $\gamma$ and $\chi_0$ takes the form, $\gamma/\chi_0 = \left(4\pi^2 k_B^2\right)/(3r)$, and assumed that the susceptibility describes a moment with $g = 1$ and $\mu_B = 1$.[42] Written explicitly, this ratio becomes, $\gamma/\chi_0 = \left(4\pi^2 k_B^2\right)/\left(3g^2 \mu_B^2 r\right)$. Substitution of this expression into **Eqn. 10** provides,

$$R_w = \frac{\pi^2 k_B^2}{3\mu_B^2} \frac{3g^2 \mu_B^2 r}{4\pi^2 k_B^2} = \frac{g^2 r}{4}. \quad (12)$$

If $g = 2$ and $r \approx 2$, as Wilson found, then $R_w \approx 2$ in the case of Kondo screening.

Another way to achieve a Wilson ratio $R_w > 1$ is when a compound is more magnetic than expected for a free-electron gas. In exchange-enhanced materials (*e.g.*, Pd, TiBe$_2$, Sc$_{0.75}$In$_{0.25}$), the measured susceptibility is given by $\chi_0 = S \chi_{Pauli,free}$, where $S = [1 - I g(E_F)]^{-1}$ is the Stoner enhancement factor and $I$ is the exchange interactions between electrons[20]. In a uniform enhancement model, the mass enhancement is given by $m^*/m = 1 + 3I\chi_{Pauli,free} \ln(1 + S/12)$, which modifies $\gamma_{free}$ ($m^*/m$) = $\gamma$. Substitution of the experimental attainable values into **Eqn.10** provides,

$$R_w = \frac{\pi^2 k_B^2}{3\mu_B^2} \frac{\chi_0}{\gamma} = \frac{\pi^2 k_B^2}{3\mu_B^2} \frac{\chi_{Pauli,free} S}{\gamma_{free} \left[1 + 3I\chi_{Pauli,free} \ln(1+S/12)\right]}. \quad (13)$$

The scaling then simplifies to,

$$R_w = \frac{S}{1 + 3Ig(E_F)\mu_B^2 \ln(1+S/12)} = \frac{S}{1+3(1-S^{-1})\ln(1+S/12)} \quad (14)$$

Using a typical value of $S = 10$, as reported for Pd,[20] the calculated ratio results in $R_w = 10 / [1+2.7\ln(1.83)] = 3.8$ (the experimental $R_w = 4.2$).

For KNi$_2$Se$_2$, $R_w = 1.7$, which aligns it with Kondo-screening and heavy-fermion behavior, despite the absence of localized magnetic behavior. Another possibility comes from the fact that effective electronic masses can diverge in the proximity of a metal-to-insulator transition, even when the transition is suppressed below $T = 0$ K, as in V$_{1-x}$Mo$_x$O$_2$ [51]. Such a mechanism is unlikely to be operable here, as local charge order should not be present at high temperatures and then disappear on cooling (the reverse would be true), as we unambiguously observe for KNi$_2$Se$_2$.

From the structural studies, we infer that the incipient CDWs originate in a band derived of Ni $d_{x^2-y^2}$ character oriented along Ni–Ni bonds, thus causing the large Ni-Ni displacements at $T = 300$ K [shown to scale in **Fig.8(d)**, and overlain with the average crystal structure in **Fig.9(a)**]; the manifold of conduction electrons arises from Ni $d_{xz}$, $d_{yz}$, and Se $4p$ states. The nearly localized charge states from these distortions are phenomenologically distinct from the impurity charge Kondo effect observed in Tl$_{1-x}$Pb$_x$Te [52]. There, Kondo behavior arises from a single-ion derived negative-*U* description motivated by disproportionation of Tl$^{2+}$ to Tl$^+$ and Tl$^{3+}$ [53]. Such a theory cannot capture the nature of bonding observed in KNi$_2$Se$_2$, where band-derived charge fluctuations from Ni–Ni interactions, rather than single-ion mixed-valence, give rise to the formation of a coherent, heavy-band.

**Conclusions**

The physical properties of KNi$_2$Se$_2$ exhibit remarkable changes on cooling: as local CDWs 'disappear', there is a significant drop in resistance, increase in carrier mobility, and increase in effective electron mass. While the thermal activation of an anharmonic phonon can explain some of these observations, additional electronic physics are required to impart the enhanced electronic mass at low temperatures, akin to the hybridization described in heavy-fermion behavior. Our results are expected to stimulate significant experimental and theoretical research into *how* cooperative Coulombic interactions and bonding give rise to emergent electronic states.

**Acknowledgments:**

This research is principally supported by the U.S. DoE, Office of Basic Energy Sciences (BES), Division of Materials Sciences and Engineering under Award DE-FG02-08ER46544. The dilution refrigerator was funded by the National Science Foundation Major Research Instrumentation Program, Grant #NSF DMR-0821005. This work benefited from the use of HIPD and NPDF at the Lujan Center at Los Alamos Neutron Science Center, funded by DoE BES. Los Alamos National Laboratory is operated by Los Alamos National Security LLC under DoE Contract DE-AC52-06NA25396. The upgrade of NPDF was funded by the National Science Foundation through grant DMR 00-76488. A portion of this work was performed at the National High Magnetic Field Laboratory, which is supported by NSF Cooperative Agreement No. DMR-0654118, the State of Florida, and the U.S. DoE. JRN acknowledges D P Shoemaker for helpful discussions regarding the RMC analysis. AVS, LW, and NPA were additionally supported by the Gordon and Betty Moore foundation and acknowledge A. Suslov for help with collection of the high-field data. The work at BNL was supported by the U.S. DoE, BES, by the Materials Sciences and Engineering Division under Contract No. DE-AC02-98CH10886 and through the use of Center for Functional Nanomaterials (CFN).

**Table I.** Debye temperatures and Sommerfield constants extracted from the heat capacity over different temperature regions.

| Low-temperature Debye Model [**Figure 1(a)**] | | | | |
|---|---|---|---|---|
| Fit region | $\beta_3$ (mJ mol$^{-1}$ K$^{-4}$) | $\beta_5$ ($\mu$J mol$^{-1}$ K$^{-6}$) | $\gamma$ (mJ mol$^{-1}$ K$^{-2}$) | |
| 2 K – 10 K | 1.03(2) [$\Theta_D$ = 211(1) K] | 9.3(2) | 44(1) | |
| Full high-temperature Double Debye Model [**Figure 1(b)**] | | | | |
| Fit region | $N$ | $S$ | $\Theta_{D1}$ (K) | $\Theta_{D2}$ (K) | $\gamma$ (mJ mol$^{-1}$ K$^{-2}$) |
| 30 K – 400 K | 5 | 2 | 327(1) K | 136(1) | 0 (fixed) |
| 30 K – 400 K | 5 | 2 | 339(1) K | 134.6(4) | 14(1) |
| 1.8 K – 400 K | 5 | 2 | 357(2) K | 128.1(3) | 22.6(1) |

**Table II.** Magnetoresistance parameters from a fit of $\rho/\rho_{18T}$ =OMR(1+[(1 + $bB^2$) / (1 + $c^2 B^2$)]) over 0 < $B$ < 18 T.

| $T$ (K) | $\mu_{Drude}$ (cm$^2$ V$^{-1}$ s$^{-1}$) | $c$ | OMR |
|---|---|---|---|
| 0.4 | 1077(2) | -4.42(8) | 0.175(2) |
| 4 | 1070(1) | -4.65(5) | 0.180(2) |
| 10 | 1026(1) | -5.63(5) | 0.193(2) |
| 15 | 896(1) | -12.9(1) | 0.217(2) |
| 20 | 798(1) | -24.6(2) | 0.254(1) |
| 30 | 626(1) | -78.9(5) | 0.328(1) |
| 35 | 587(1) | -86(1) | 0.425(3) |
| 40 | 527(1) | -155(1) | 0.448(2) |
| 50 | 385(1) | -481(15) | 0.623(4) |
| 60 | 417(1) | -411(5) | 0.558(2) |
| 65 | 387(1) | -634(12) | 0.581(3) |

**Table III.** Crystal Structure parameters of KNi$_2$Se$_2$ obtained from Rietveld Analysis of NPD data. Space group *I*4/*mmm*. K: 2*a*, (0, 0, 0); Ni: 4*d*, (0, 0.5, 0.25); Se: 4*e*, (0, 0, *z*).

| Parameter | 15 K | 300 K |
|---|---|---|
| $a$ | 3.89295(2) | 3.90979(2) |
| $c$ | 13.3158(1) | 13.4145(2) |
| $z$ (Se) | 0.35469(3) | 0.35388(5) |
| $U_{11}$(K) | 0.0044(1) | 0.0145(5) |
| $U_{33}$(K) | 0.0093(4) | 0.027(1) |
| $U_{11}$(Ni) | 0.00431(4) | 0.0146(2) |
| $U_{33}$(Ni) | 0.01176(9) | 0.0328(4) |
| $U_{11}$(Se) | 0.00359(5) | 0.0117(2) |
| $U_{33}$(Se) | 0.0084(1) | 0.0226(5) |